\documentclass[aps,prd,showpacs,nobibnotes,twocolumn,
nobalancelastpage,superscriptaddress,preprintnumbers,
nofootinbib]{revtex4}
\usepackage{graphicx}
\usepackage{amsmath,amssymb}
\DeclareMathAlphabet{\mathpzc}{OT1}{pzc}{m}{it}

\begin{document}


\title{Towards Closing the Window on Strongly Interacting Dark Matter:\\
 Far-Reaching Constraints from Earth's Heat Flow}
\author{Gregory D. Mack}
\affiliation{Department of Physics, Ohio State University,
Columbus, Ohio 43210}
\affiliation{Center for Cosmology and Astro-Particle Physics,
Ohio State University, Columbus, Ohio 43210}

\author{John F. Beacom}
\affiliation{Department of Physics, Ohio State University,
Columbus, Ohio 43210}
\affiliation{Center for Cosmology and Astro-Particle Physics,
Ohio State University, Columbus, Ohio 43210}
\affiliation{Department of Astronomy, Ohio State University,
Columbus, Ohio 43210}

\author{Gianfranco Bertone}
\affiliation{Institut d'Astrophysique de Paris, UMR 7095-CNRS,
Universit\'e Pierre et Marie Curie, 98bis boulevard Arago, 75014 Paris,
France \\
{\tt gdmack@mps.ohio-state.edu, beacom@mps.ohio-state.edu,
bertone@iap.fr}}

\begin{abstract}
We point out a new and largely model-independent constraint on the
dark matter scattering cross section with nucleons, applying when this
quantity is larger than for typical weakly interacting dark matter
candidates.  When the dark matter capture rate in Earth is efficient,
the rate of energy deposition by dark matter self-annihilation
products would grossly exceed the measured heat flow of Earth.  This
improves the spin-independent cross section constraints by many orders
of magnitude, and closes the window between astrophysical constraints
(at very large cross sections) and underground detector constraints
(at small cross sections).  In the applicable mass range, from $\sim$
1 to $\sim$ 10$^{10}$ GeV, the scattering cross section of dark matter
with nucleons is then bounded from {\it above} by the latter
constraints, and hence must be truly weak, as usually assumed.
\end{abstract}

\date{30 May 2007}



\maketitle

\section{Introduction}
There is a large body of evidence for the existence of dark matter,
but its basic properties -- especially its mass and scattering cross
section with nucleons -- remain unknown.  Assuming dark matter is a
thermal relic of the early universe, weakly interacting massive
particles are prime candidates, suggested by constraints on the dark
matter mass and self-annihilation cross section from the present
average mass density~\cite{Bertone}.  However, as this remains
unproven, it is important to systematically test the properties of
dark matter particles using only late-universe constraints.  In 1990,
Starkman, Gould, Esmailzadeh, and Dimopoulos~\cite{Starkman:1990nj}
examined the possibility of strongly interacting dark matter, noting
that it indeed had not been ruled out.  Many authors since have
explored further constraints and candidates.  In this literature,
``strongly interacting'' denotes cross sections significantly larger
than those of the weak interactions; it does not necessarily mean via
the usual strong interactions between hadrons.  We generally consider
the constraints in the plane of dark matter mass $m_\chi$ and
spin-independent scattering cross section with nucleons $\sigma_{\chi
N}$.

\begin{figure}[!ht]
\includegraphics[width=3.25in,clip=true]{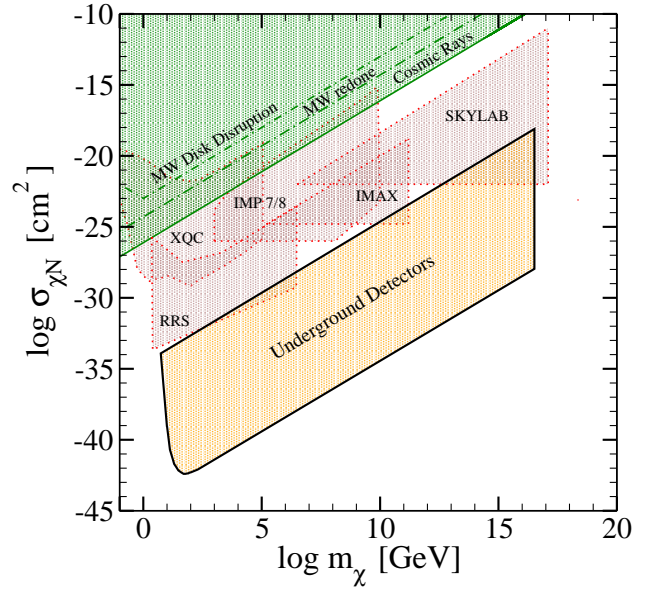}
\caption{\label{regions} Excluded regions in the
$\sigma_{\chi N}$--$m_\chi$ plane, not yet including the results of
this paper.  From top to bottom, these come from astrophysical
constraints
(dark-shaded)~\cite{Starkman:1990nj,Natarajan:2002cw,Chen:2002yh,Cyburt:2002uw},
re-analyses of high-altitude detectors
(medium-shaded)~\cite{Rich:1987st,Starkman:1990nj,Wandelt,Erickcek:2007jv},
and underground direct dark matter detectors
(light-shaded)~\cite{Akerib:2006ri,Albuquerque:2003ei,Sanglard:2005we,Bernabei:1999ui}.
The dark matter number density scales as 1/$m_\chi$, and the
scattering rates as $\sigma_{\chi N}$/$m_\chi$; for a fixed scattering
rate, the required cross section then scales as $m_\chi$.  We will
develop a constraint from Earth heating by dark matter annihilation to
more definitively exclude the window between the astrophysical and
underground constraints.}
\end{figure}

Figure~\ref{regions} summarizes astrophysical, high-altitude
balloon/rocket/satellite detector, and underground detector
constraints in the $\sigma_{\chi N}$--$m_\chi$ plane.  Astrophysical
limits such as the stability of the Milky Way disk constrain very
large cross sections~\cite{Starkman:1990nj,Natarajan:2002cw}.
Accompanying and comparable limits include those from cosmic rays and
the cosmic microwave background~\cite{Chen:2002yh,Cyburt:2002uw}.
Small cross sections are probed by CDMS and other underground
detectors~\cite{Akerib:2006ri,Albuquerque:2003ei,Sanglard:2005we,Bernabei:1999ui}.
A dark matter (DM) particle can be directly detected if $\sigma_{\chi
N}$ is strong enough to cause a nuclear recoil in the detector, but
only if it is weak enough to allow the
DM to pass through Earth to the
detector.

In between the astrophysical and underground limits is the window in
which $\sigma_{\chi N}$ can be relatively
large~\cite{Starkman:1990nj}.  High-altitude detectors in and above
the atmosphere have been used to exclude moderate-to-strong values of
the cross section in this region
\cite{Rich:1987st,Starkman:1990nj,Wandelt,Erickcek:2007jv}.  However,
there are still large gaps not excluded.  There also is some doubt
associated with these exclusions, as some of the experiments were not
specifically designed to look for DM, nor were they always analyzed
for this purpose by people associated with the projects.  In fact, the
exclusion from the X-ray Quantum Calorimetry experiment was recently
reanalyzed~\cite{Erickcek:2007jv} and it changed substantially from
earlier estimates~\cite{Wandelt}.  If this intermediate region can be
closed, then underground detectors would set the upper limit on
$\sigma_{\chi N}$.  That would mean that these detectors are generally
looking in the right cross section range and that DM-nucleon
scattering interactions are indeed totally irrelevant in astrophysics.

We investigate cross sections between the astrophysical and
underground limits, and show that $\sigma_{\chi N}$ is large enough
for Earth to efficiently capture DM.  Incoming DM will scatter off
nucleons, lose energy, and become gravitationally captured once below
Earth's escape velocity (Section 4).  If this capture is maximally
efficient, the rate is $2 \times 10^{25}
\,(\textrm{GeV}/m_\chi)$ s$^{-1}$.  
The gravitationally-captured DM will drift to the bottom of the potential
well, Earth's core.  Self-annihilation results if the DM is its
own antiparticle, and we assume Standard Model final state particles
so that these products will deposit nearly all their energy in the
core.  

Inside a region in the $\sigma_{\chi N}$--$m_\chi$ plane that will be
defined, too much heat would be produced relative to the actual
measured value of Earth's heat flow.  The maximal heating rate
obtained via macroscopic considerations is $\simeq$ 3330 TeraWatts
(TW), and follows the maximal capture rate, assuming that Earth is
opaque with a geometric cross section.  Note that the flux of DM
scales as $1/m_\chi$, while the heat energy from annihilations scales
as $m_\chi$, {\it yielding a heat flow that is independent of DM
mass}.  The efficient capture we consider leads to a very similar
heating rate, though it is based on a realistic calculation of
microscopic DM-nucleon scattering, as discussed below.  DM
interactions with Earth have been previously studied in great detail,
e.g.,
Refs~\cite{Starkman:1990nj,Gaisser:1986ha,Gould1,Gould2,Gould3,Gould4,
Gould5,Gould6,Gould7,KSW}, but those investigations
generally considered only weak cross sections for which capture
is inefficient.

In our analysis, the $\sigma_{\chi N}$ exclusion region arises from
the captured DM's self-annihilation energy exceeding Earth's internal
heat flow.  This region is limited below by the efficient capture of
DM (Section 4), and above by being weak enough to allow sufficient
time for the DM to drift to the core (Section 5).  These two limits
define the region in which DM heating occurs.  Why is it important?
Earth's received solar energy is large, about $170,000$ TW
\cite{EarthSunbudget}, but it
is all reflected or re-radiated.  The internal heat flow is much less,
about 44 TW (Section 3)~\cite{TextBook}.  Inside this bounded region
for $\sigma_{\chi N}$, DM heating would exceed the measured rate by
about two orders of magnitude, and therefore is not allowed.  We will
show that this appears to close the window noted above in
Fig.~\ref{regions}, up to about $m_\chi \simeq 10^{10}$ GeV.  In order
to be certain of this, however, we call for new analyses of the
aforementioned constraints, especially the exact region excluded by
CDMS and other underground detectors.  Our emphasis is not on further
debate of the details of specific open gaps, but rather on providing a
new and independent constraint.  In Table I, we summarize the heat
values relevant to this paper.  While the origin of Earth's heat flow
is not completely understood, we emphasize that we are not trying to
account for any portion of it with heating from DM.

\begin{table}
\caption{Relevant heat flow values.  The top entries are measured, while
the lower entries are the calculated potential effects of dark
matter.}
\begin{center}
  \begin{tabular}{|l|r|}
    \hline
     Heat Source & Heating Rate\\
    \hline\hline
    Solar (received and returned) & 170,000 TW\\
    
    Internal (measured) & 44.2 $\pm$ 1 TW\\
    \hline
    DM annihilation (opaque Earth) & 3330 TW\\
    
    DM annihilation (our assumptions) & 3260 TW\\

    DM kinetic heating & $\sim$ 3000 $\times$ 10$^{-6}$ TW\\
    \hline
  \end{tabular}
\end{center}
\end{table}

There has been some previous work on the heating of planets by DM
annihilation
~\cite{KSW,Fukugita:1988,Dimopoulos:1989hk,Kawasaki:1991eu,Abbas:1996kk,Uranus,Farrar:2005zd,Zaharijas:2004jv}.
These papers have mostly focused on the Jovian planets, for which the
internal heat flow values are deduced from their infrared
radiation~\cite{AstroBook}.  In some
cases~\cite{KSW,Fukugita:1988,Dimopoulos:1989hk,Kawasaki:1991eu,Abbas:1996kk},
DM annihilation was invoked to explain the anomalously large heat flow
values of Jupiter and Saturn, while in other
cases~\cite{Uranus,Farrar:2005zd,Zaharijas:2004jv}, the low heat flow
value of Uranus was used to constrain DM annihilation. An additional
reason for the focus on these large planets is that they will be able
to stop DM particles of smaller cross section than Earth can
(Ref.~\cite{Abbas:1996kk} considered Earth, but invoked an extreme DM
clumping factor to overcome the weakly interacting cross section).
However, as we argue below in Section 6, the more relevant criterion
is how significant of an excess heat flow could be produced by DM
annihilation, and this is much more favorable for Earth.  (If this
criterion is met, then the ranges of excluded cross sections will
simply shift for different planets.)  Furthermore, the detailed
knowledge of Earth's properties gives much more robust results.  In
this paper, we are presenting the first detailed and systematic study
of the broad exclusion region in the $\sigma_{\chi N}$--$m_\chi$ plane
that is based on not overheating Earth.

Our constraints depend on DM being its own antiparticle, so that
annihilation may occur (or, if it is not, that the DM-antiDM asymmetry
not be too large).  This is a mild and common assumption.  The heating
due to kinetic energy transfer is negligible.  Since the DM velocity
is $\simeq$ 10$^{-3}c$, kinetic heating is $\sim$ 10$^{-6}$ that from
annihilation, and would provide no constraint (Section 4).  The
model-independent nature of our annihilation constraints arises from
the nearly complete insensitivity to which Standard Model particles
are produced in the DM annihilations, and at what energies.  All final
states except neutrinos will deposit all of their energy in Earth's
core.  (Above about 100 TeV, neutrinos will, too.)  Since the possible
heating rate ($>$ 3000 TW) is so large compared to the measured rate
($\sim$ 40 TW), in effect we only require that not more than
$\sim$99\% of the energy goes into low-energy neutrinos, which is an
extremely modest assumption.

Some of the annihilation products will likely be neutrinos, and these
may initiate signals in neutrino detectors, e.g., as upward-going
muons~\cite{Starkman:1990nj,AHK,Crotty,Albuquerque:2002bj,Desai:2004pq,
Bottino:1994xp,Achterberg:2006jf,Cirelli:2005gh,Bergstrom:1997tp}.
While the derived cross section limits can be constraining, they
strongly depend on the branching ratio to neutrinos and the neutrino
energies.  Comprehensive constraints based on neutrino fluxes for the
full range of DM masses appear to be unavailable; most papers have
concentrated on the 1--1000 GeV range, and a few have considered
masses above $10^8$ GeV.  We note that the constraints for DM masses
above about $10^{10}$ GeV may require annihilation cross sections
above the unitarity bound, as discussed below.  As this paper is meant
to be a model-independent, direct approach to DM properties based on
the DM density alone, we do not include these neutrino constraints.

We review the current DM constraints in Section 2, review Earth's heat
flow in Section 3, calculate the DM capture, annihilation, and heating
rates in Sections 4 and 5, and close with discussions and conclusions
in Section 6.

\section{Review of Prior Constraints}
Figure~\ref{regions} shows the current constraints in the
$\sigma_{\chi N}$--$m_\chi$ plane.  As we will show, the derived
exclusion region found by the requirement of not overheating Earth
using DM annihilation lies in the uncertain intermediate area between
the astrophysical and underground constraints.

\subsection{Indirect Astrophysical Constraints}

If $\sigma_{\chi N}$ were too large, DM particles in a galactic halo
would scatter too frequently with the baryonic disk of a spiral
galaxy, and would significantly disrupt it.  Using the integrity of
the Milky Way disk, Starkman et al.~\cite{Starkman:1990nj} restrict
the cross section to $\sigma_{\chi N} < 5 \times 10^{-24}
({m_\chi}/{\textrm{GeV}})$ cm$^2$. A more detailed study by Natarajan
et al.~\cite{Natarajan:2002cw} requires $\sigma_{\chi N} < 5 \times
10^{-25} ({m_\chi}/{\textrm{GeV}})$ cm$^2$.  Both of these limits
consider DM scattering only with hydrogen.  As shown below in
Eq.~(\ref{Ascaling}), the spin-independent DM-nucleon cross section
scales as $A^4$ for large $m_\chi$, and though the number density of
helium ($A =4$) is about 10 times less than that of hydrogen ($A =
1$), taking it into account could improve these constraints by
$\simeq$ 256/10 $\simeq$ 25.  Chivukula et al.~\cite{Chivukula:1989cc}
showed that charged dark matter could be limited through its ionizing
effects on interstellar clouds; this technique could be adapted for
strongly interacting dark matter.

Strong scattering of DM and baryons would also affect the cosmic
microwave background radiation.  Adding stronger DM-baryon
interactions increases the viscosity of the baryon-photon
fluid~\cite{Chen:2002yh}.  A strong coupling of baryons and DM would
generate denser clumps of gravitationally-interacting matter, and the
photons would not be able to push them as far apart.  The peaks in the
cosmic microwave background power spectrum would be damped, with the
exception of the first one.  The resulting constraint is $\sigma_{\chi
N} < 3 \times 10^{-24} ({m_\chi}/{\textrm{GeV}})$
cm$^2$~\cite{Chen:2002yh}, and is not shown in Fig.~\ref{regions}.
These results do take helium into account, but do so only using $A^2$
instead of $A^4$.  This possible change, along with the much more
precise cosmic microwave background radiation data available
currently, calls for a detailed re-analysis of this limit, which
should strengthen it.

Cosmic ray protons interact inelastically with interstellar protons,
breaking the protons and creating neutral pions that decay to
high-energy gamma rays.  A similar situation could occur with a cosmic
ray beam on DM targets instead~\cite{Cyburt:2002uw}. The fundamental
interaction is between the quarks in the nucleon and the DM; it is
very unlikely that all quarks will be struck equally, and the
subsequent destruction of the nucleon creates pions.  If the
DM-nucleon cross section were high enough, the resulting gamma rays
would be readily detectable.  From this, Cyburt et
al.~\cite{Cyburt:2002uw} place an upper limit of $\sigma_{\chi N} <
7.6 \times 10^{-27} ({m_\chi}/{\textrm{GeV}})$ cm$^2$.  Improvements
could probably be made easily with a more realistic treatment of the
gamma-ray data.

\subsection{Direct Detection Constraints}
Underground detector experiments have played a large role in limiting
DM that can elastically scatter nuclei, giving the nuclei small but
measurable kinetic energies.  Due to the cosmic ray background, this
type of detector is located underground.  The usual weakly interacting
DM candidates easily pass through the atmosphere and Earth en route to
the detector.  However, for large $\sigma_{\chi N}$ the DM would lose
energy through scattering before reaching the detector, decreasing
detection rates.

Albuquerque and Baudis~\cite{Albuquerque:2003ei} have explored
constraints at relatively large cross sections and large masses using
results from CDMS and EDELWEISS.  In Fig.~\ref{regions}, we present a
crude estimate of the current underground detector exclusion region.
The top line is defined by the ability of a DM particle to make it
through the atmosphere~\cite{HarryNelson} and Earth to the detector
without losing too much energy \cite{Albuquerque:2003ei}.  The lower
left corner and nearby points are taken from the official CDMS
papers~\cite{Akerib:2006ri} with the aid of their
website~\cite{CDMSweb}.  The right edge is taken from
DAMA~\cite{Bernabei:1999ui}.  As the mass of the DM increases, the
number density (and hence the flux through Earth) decreases.  At the
largest $m_\chi$ values, the scattering rate within a finite time
vanishes.  Finally, we have extrapolated each of these constraints to
meet each other, connecting them consistently.  We call for a complete
and official analysis of the exact region that CDMS and other direct
detectors exclude.  Our focus is on the cross sections in between the
underground detectors and astrophysical limits.

To investigate cross sections in this middle range, direct detectors
must be situated above Earth's atmosphere, in high-altitude balloons,
rockets, or satellites.  Several such detectors have been analyzed for
this purpose, though they were not all originally intended to study
DM.  Since these large $\sigma_{\chi N}$ limits have in some cases
been calculated by people not connected with the original experiments,
some caution is required.  Nevertheless, in Fig.~\ref{regions} we show
the claimed exclusion regions, following Starkman et
al.~\cite{Starkman:1990nj} and Rich et al.~\cite{Rich:1987st}, along
with Wandelt et al.~\cite{Wandelt} and Erickcek et
al.~\cite{Erickcek:2007jv} (including the primary
references~\cite{McGuire:1994pq,Shirk:1978,Snowden-Ifft:1990,Anderson:1995}).
We are primarily in accordance with Erickcek et al.  These regions
span masses of almost 0.1 GeV to 10$^{16}$ GeV, and cross sections
between roughly 10$^{-33}$ cm$^2$ and 10$^{-11}$ cm$^2$.  These
include the Pioneer 11 spacecraft and Skylab, the IMP 7/8 cosmic ray
silicon detector satellite, the X-ray Quantum Calorimetry experiment
(XQC), and the balloon-borne IMAX.  These regions are likely ruled
out, but not in absolute certainty, and there are gaps between them.
The Pioneer 11 region is completely covered by the IMP 7/8 and XQC
regions, and is therefore not shown in Fig.~\ref{regions}.  The region
labeled RRS is Rich et al.'s analysis of a silicon semiconductor
detector near the top of the atmosphere, truncated according to
Starkman et al., and adjusted with the appropriate $A$-scaling as in
Eq.~(\ref{Ascaling}).

\section{Earth's Heat Flow}
Heat from the Sun warms Earth, but it is not retained.  If all the
incident sunlight were absorbed by Earth, the heating rate would be
about 170,000 TW~\cite{EarthSunbudget}.  Some of it is reflected by
the atmosphere, clouds, and surface, and the rest is absorbed at
depths very close to the surface and then
re-radiated~\cite{AstroBook}.  Earth's blackbody temperature would be
about 280 K, and it is observed to be between 250 and 300 K,
supporting the idea of Earth-Sun heat equilibrium.  Internal heating
therefore has minimal effects on the overall heat of
Earth~\cite{AstroBook}.

Our focus is on this {\it internal} heat flow of Earth, as measured
underground.  Geologists have extensively studied Earth's internal
heat for decades~\cite{Pollack}.  To make a measurement, a borehole is
drilled kilometers deep into the ground.  The temperature gradient in
that borehole is recorded, and that quantity multiplied by the thermal
conductivity of the relevant material yields a heat flux
\cite{Pollack,BlackBook}.

The deepest borehole is about 12 kilometers, which is still rather
close to Earth's surface.  Typical temperature gradients are between
10 and 50 K/km, but these cannot hold for lower depths.  If they did,
all rock in the deeper parts of Earth would be molten, in
contradiction to seismic measurements, which show that shear waves can
propagate through the mantle~\cite{BlackBook}.  Current estimates
place temperature gradients deep inside Earth between 0.6 and 0.8 K/km
\cite{BlackBook}.

More than 20,000 borehole measurements have been made over Earth's
surface.  Averaging over the continents and oceans, there is a heat
flux of 0.087 $\pm$ 0.002 W/m$^2$~\cite{Pollack,TextBook}.
Integrating this flux over the surface of Earth gives a heat flow of
44.2 $\pm$ 1 TW \cite{Pollack,TextBook}.  Again, the heat flux is
directly measured underground, all over Earth, and is independent of
the solar flux, Earth's atmosphere, and anything else above Earth's
surface.  Obviously, the possibility to make direct heat flow
measurements under the surface is unique to Earth.

While the heat flow value is known well, the origin of the heat is
not, and in fact is undergoing much theoretical debate
\cite{vFrese,Panero}.  Some specific contributors are known, however.  
The decay of radioactive elements produces a significant amount;
uranium and thorium decay in the crust generates about forty percent
of the total~\cite{TextBook}.  Potassium adds to this, though there is
much less of it in the crust.  However, there is potentially a large
amount in the mantle and perhaps even the outer core \cite{TextBook}.
KamLAND has a hint of detected neutrinos coming from uranium and
thorium decays~\cite{Araki:2005qa}, and it (along with other
detectors) could potentially help to make the heat contribution from
them more
accurate~\cite{Fiorentini:2005ma,Fiorentini1,Fields,Krauss:1983zn,
Hochmuth:2005nh,Dye:2006gx,Fogli:2006zm,Miramonti:2006kp,Rothschild:1997dd,Raghavan:1997gw}.
Larger concentrations of uranium and thorium are excluded by KamLAND,
and theoretical predictions from the Bulk Silicate Earth model are
consistent with the forty percent value
\cite{Fiorentini:2005ma,Fiorentini1,Fields}.  The remaining heat is due to processes
in the core and perhaps even the mantle, although specific knowledge
of Earth's interior is limited \cite{OrangeBook}.

The residual heat flow, which we assume to be 20
TW~\cite{vFrese,Panero}, we use as the target limit for the heat flow
from DM annihilation.  Models give values of the core's heat output
between 2.3 TW and 21 TW, supporting the conservative choice of 20 TW
\cite{OrangeBook}.  Annihilation scenarios creating heat flows greater
than 20 TW are therefore excluded.  In fact, if heating by DM
annihilation is important at all, we show that it typically would
exceed this value by more than two orders of magnitude.  It is
important to note that we are not trying to solve geological heat
problems with DM, and in fact our analysis implies it is very unlikely
that DM is contributing to Earth's internal heat flow, which is
interesting in itself.

\section{Dark Matter Capture Rate of Earth}
The DM mass density, $\rho_\chi = n_\chi m_\chi$, in the neighborhood
of the solar system is about 0.3 GeV/cm$^3$~\cite{Bertone}.  Neither
the mass nor the number density are separately known.  The DM is
believed to follow a nonrelativistic Maxwell-Boltzmann velocity
distribution with an average speed of about 270 km/s.  If a DM
particle scatters a sufficient number of times while passing through
Earth, its speed will fall below the surface escape speed, 11.2 km/s.
Having therefore been gravitationally captured, it will then orbit the
center of Earth, losing energy with each subsequent scattering until
it settles into a thermal distribution in equilibrium with the nuclei
in the core.  For the usual weak cross sections Earth is effectively
transparent, and scattering and capture are very inefficient.  In
contrast, we will consider only large cross sections for which capture
is almost fully efficient.  Note that for our purposes, the scattering
history is irrelevant as long as capture occurs; in particular, the
depth in the atmosphere or Earth of the first scattering has no
bearing on the results.  The energies of the individual struck nuclei
are also irrelevant, unlike in direct detection experiments.  We just
require that the DM is captured and ultimately annihilated.

\subsection{Maximum Capture Rate}
We begin by considering the maximum possible capture rate of DM in
Earth, which corresponds to Earth being totally opaque.  Although our
final calculations will involve the microscopic scattering cross
section of DM on nuclei, this initial example deals with just the
macroscopic geometric cross section of Earth.  The flux per solid
angle of DM near Earth is $n_\chi v_\chi$/4$\pi$, where $n_\chi$ is
the DM number density, and $v_\chi$ is the average DM velocity.  Since
Earth is taken to be opaque, the solid angle acceptance at each point
on the surface is 2$\pi$ sr.  Thus the flux at Earth's surface is
$n_\chi v_\chi$/2.  The capture rate is then found by multiplying by
Earth's geometric cross section, $\sigma_\oplus~=$ 4$\pi R_\oplus^2
\simeq 5.1
\times 10^{18}$ cm$^2$.  Since $n_\chi$ is not known, this is
$(\rho_\chi/m_\chi)\sigma_\oplus v_\chi$.  For $v_\chi$ = 270 km/s,
this maximal capture rate is
\begin{equation}
\Gamma_{C}^{max} = 2 \times
10^{25}\left(\frac{\textrm{GeV}}{m_\chi}\right)\textrm{s}^{-1}.
\label{maxcapture}
\end{equation}
We will show that our results depend only logarithmically on the DM
velocity, and hence are insensitive to the details of the velocity
distribution.

This maximal capture rate estimate is too simplistic, as it assumes
that merely coming into contact with Earth, interacting with any
thickness, will result in DM capture.  Instead, we define opaqueness
to be limited to path lengths greater than 0.2 $R_\oplus$, a value
that incorporates the largest 90\% of path lengths through Earth.
This reduces the capture rate, but only by about 2\%.  We therefore
adopt 0.2 $R_\oplus$ as our minimum thickness to determine efficient
scattering.  This length, translated into a chord going through the
spherical Earth, defines the new effective area for Earth.  The
midpoint of the chord lies at a distance of 0.99$R_\oplus$ from
Earth's center.  Thus, practically speaking, nearly all DM passing
through Earth will encounter sufficient material.  The above
requirements exclude glancing trajectories from consideration, for
which there would be some probability of reflection from the
atmosphere~\cite{Uranus,Zaharijas:2004jv}; note also that the
exclusion region in Section 4 would be unaffected by taking this into
account, since the DM heating of Earth would still be excessive.

The type of nucleus with which DM scatters depends on its initial
trajectory through Earth.  For a minimum path length of 0.2
$R_\oplus$, this trajectory runs through the crust, where the density
is 3.6 g/cm$^3$ \cite{Density}, and the most abundant element is
oxygen \cite{Gould2}.  Choosing this path length and density are
conservative steps.  Any larger path length would result in more
efficient capture, and a higher density and heavier composition
(corresponding to a larger chord and therefore a different target
nucleus, such as iron, which is the most abundant element in the core)
would as well.  A more complex crust or mantle composition, such as
30\% oxygen, 15\% silicon, 14\% magnesium and smaller contributions
from other elements~\cite{Gould2}, would stop DM $\sim$2 times more
effectively.

\subsection{Dark Matter Scattering on Nuclei}
When a DM particle (at $v_\chi~\simeq$ 10$^{-3}c$) elastically
scatters with a nucleus/nucleon (at rest) in Earth, it decreases in
energy and velocity.  After one scattering with a nucleus of mass
$m_A$, DM with mass $m_\chi$ and initial velocity $v_i$ will have a
new velocity of
\begin{eqnarray}
\frac{v_f}{v_i}~&=&~ \!\!\!
\sqrt{1 -
    2\frac{m_A  m_\chi}{(m_\chi + m_A)^2}(1-\cos\theta_{cm})}~,
\label{velocloss}\\
&\underset{m_\chi \gg m_A}{\longrightarrow}& 
\sqrt{1-2\frac{m_A}{m_\chi}(1-\cos\theta_{cm})}.
\end{eqnarray}
All quantities are in the lab frame, except the recoil angle,
$\theta_{cm}$, which is most usefully defined in the center of mass
frame (see Landau and Lifschitz~\cite{LandL}).  Here and below we give
the large $m_\chi$ limit for demonstration purposes, but we use the
full forms of the equations for our results.  After scattering, the DM
has a new kinetic energy,
\begin{eqnarray}
&KE_f^{~\chi}& = \nonumber \\
&&\frac{1}{2} m_\chi v_i^2\left(1 -
    2\frac{m_A m_\chi}{(m_\chi + m_A)^2}(1-\cos\theta_{cm})\right)
\label{KE}\\
&\underset{m_\chi \gg m_A}{\longrightarrow}& KE_i^{~\chi}\left(1-
2\frac{m_A}{m_\chi}(1-\cos\theta_{cm})\right).
\end{eqnarray}
The nucleus then obtains a kinetic energy of
\begin{eqnarray}
\hspace{-0.5cm}&&KE^{A} = KE_{i}^{~\chi} - KE_{f}^{~\chi} \nonumber \\ 
&&=\frac{1}{2}m_\chi v_i^2\left(1 - 1 +
    2\frac{m_A m_\chi}{(m_\chi + m_A)^2}(1-\cos\theta_{cm})\right) \\
&&\underset{m_\chi \gg m_A}{\longrightarrow} KE_{i}^{~\chi}~
2\frac{m_A}{m_\chi}(1-\cos\theta_{cm})\\
&&= m_Av_i^2(1-\cos\theta_{cm})
\end{eqnarray}
From the kinetic energy, the momentum transfer in the large $m_\chi$
limit is:
\begin{eqnarray}
KE &=& \frac{|\vec{q}|^2}{2 m_A} = m_A v_i^2 (1-\cos{\theta_{cm}})\\
|\vec{q}|^2 &=& 2(m_A v_i)^2(1-\cos{\theta_{cm}}).
\end{eqnarray}

In order to maintain consistency with others, we work with $n$ and
$\sigma$ in \it{nucleon~}\rm units even though the target we choose
(oxygen) is a nucleus.  This means that $n_A$ (where $A$ represents
the mass number of the target) is
\begin{equation}
n_A = \frac{n}{A} = \frac{\rho}{m_N A}.
\end{equation}
In turn, the cross section for spin-independent s-wave elastic
scattering is represented as
\begin{eqnarray}
\sigma_{\chi A} &=& A^2
  \left(\frac{\mu(A)}{\mu(N)}\right)^2 \sigma_{\chi N}~ 
\label{Ascaling}
\\
&\underset{m_\chi \gg m_A}{\longrightarrow}& A^4 \sigma_{\chi N}.\nonumber
\end{eqnarray}
Here $A$ is the mass number of the target nucleus, which equals
$m_A/m_N$, and $\mu(A\textrm{ or }N)$ is the reduced mass of the DM
particle and the target.

The $A^2$ factor arises because at these low momentum transfers, the
nucleus is not resolved and the DM is assumed to couple coherently to
the net ``charge'' -- the number of nucleons.  (If this coherence is
somehow lost, a factor $A$ would still remain for incoherent
scattering.)  The momentum transfer $q = \sqrt{2 m_A KE^{A}} \simeq
m_A v_i$ corresponds to a length scale of $\simeq$ 10 fm for oxygen,
much larger than the nucleus.  We find that the corresponding nuclear
form factor when the DM mass is comparable to the target mass is
$\simeq$ 0.99.  The square of the reduced mass ratio arises from the
Born approximation for scattering, which is based on the two-particle
Schr\"{o}dinger equation cast as a single particle with relative
coordinates and reduced mass~\cite{Bethe}.  The spin-dependent
scattering cross section does not have the $A^2$ factor in
Eq.~(\ref{Ascaling})~\cite{Starkman:1990nj}.  Our constraints could be
scaled to represent this case by also taking into account the relative
abundance of target nuclei with nonzero spin in Earth, which is of
order 1\%.

Note that if $m_\chi = m_A$, and $\theta_{cm} = \pi$, the DM can
transfer all of its momentum to the struck nucleus, losing all of its
energy in a single scattering through this scattering
resonance~\cite{Gould2}.  Taking this into account would make our
constraints stronger over a small range of masses, but we neglect it.
The nuclear recoil energy from this resonance is then
$\frac{1}{2}m_\chi v_i^2$.  Since $v_i$ is on average 270 km/s, this
means that the maximum energy transferred from a collision is $\sim$
10$^{-6}$ that of the annihilation energy, $m_\chi c^2$.

\subsection{Dark Matter Capture Efficiency}

From the full or approximate form of Eq.~(\ref{KE}), we see that the
DM kinetic energy is decreased by a multiplicative factor that is
linear in $\cos{\theta_{\rm cm}}$.  If, in each independent
scattering, we average over $\cos{\theta_{\rm cm}}$, the average
factor by which the kinetic energy is reduced in one or many
scatterings will simply be that obtained by setting $\cos{\theta_{\rm
cm}}$ = 0 throughout.  (For s-wave scattering, the $\cos{\theta_{\rm
cm}}$ distribution is uniform.)

We will define efficient capture so that the heating is maximized.  To
be gravitationally trapped, a DM particle must be below the escape
speed of Earth ($v_{\rm esc}$ = 11.2 km/s), or equivalently, its
kinetic energy must be less than $\frac{1}{2} m_\chi v^2_{\rm esc}$.
After one scattering event, the DM kinetic energy is reduced:
\begin{equation}
KE_f^{~\chi} = KE_i^{~\chi}\,f(m_\chi).
\end{equation}
In successive collisions, this is compounded until 
\begin{equation}
\frac{1}{2} m_\chi v_{\rm esc}^{~2}= \frac{1}{2}
m_\chi v_i^{~2}\, [f(m_\chi)]^{N_{\rm scat}}.
\label{compounded}
\end{equation}
Note that for collinear scatterings, the velocity loss in
Eq,~(\ref{velocloss}) is also speed loss, leading to the same
definition of $N_{\rm scat}$.

Therefore, on average, the number of scatterings required to
gravitationally capture the DM is
\begin{eqnarray} 
N_{\rm scat} &=&\frac{-2~\ln\left({v_i}/{v_{\rm esc}}\right)}{\ln\left[1 - 
    2\frac{m_A  m_\chi} {\left(m_\chi + m_A\right)^{2}}
	\right]}\label{scatter}\\
\underset{m_\chi \gg m_A}{\longrightarrow}&&
~\frac{m_\chi}{m_A}\,{\ln\left({v_i}/{v_\textrm{esc}}\right)},
\label{stopping}
\end{eqnarray}
where we have set $\cos{\theta_{\rm cm}}$ = 0, since this corresponds
to the average fractional change in the kinetic energy.  Again, for
simplicity the same element is taken to be the target each time.  Note
that since the initial DM velocity is inside the logarithm, $N_{\rm
scat}$ is insensitive to even large changes in the assumed initial
velocity.

The number of scatterings for a given mass is large. A DM particle
that has the same mass as the target nucleus will scatter about 10
times before it is captured.  Note that the required $N_{{\rm scat}}$
scales as $m_\chi$ in the large mass limit, becoming very large: for
$m_\chi$ above 16 TeV (10$^3$ times the target mass), $N_{\rm scat}$
is already larger than 3000.  The actual energy losses in individual
collisions are irrelevant for our analysis, as we require only that
the DM is captured after many collisions.  For large values of $N_{\rm
scat}$, all scattering histories will be well-characterized by the
average case.

So far, these equations have just been kinematics; the required
$N_{\rm scat}$ for stopping has not yet been made specific to Earth.
It becomes Earth-specific by relating $N_{\rm scat}$ to the path
length in Earth $L$ and the mean free path $\lambda$:
\begin{equation}
N_{\rm scat} = \frac{L}{\lambda} = L n_A \sigma_{\chi A}.
\label{N}
\end{equation}
The column density of Earth then defines the required cross section to
generate $N_{{\rm scat}}$ scatterings.  The shortest path the particle
could travel in is a straight line, so we use that as the minimum.
Any other path would be longer, and hence more effective at capture.
This therefore defines the most conservative limit on $\sigma_{\chi
A}$.  Since we have fixed $\cos{\theta_{\rm cm}}$ to be 0 on average,
in fact the path will not be completely straight.  However, the lab
frame scattering angles are small.

For elastic collisions between two particles, the range of scattering
angles in the lab frame depends on the two masses, $m_1$ and $m_2$.
There is a maximum scattering angle when one mass is initially at rest
in the lab frame (in this case, $m_2$)~\cite{LandL}.  If $m_1~<~m_2$,
there is no restriction on the scattering angle, which is defined in
relation to $m_1$'s initial direction ($m_2$ is at rest).  However, if
$m_1~>~m_2$, then
\begin{equation}
\sin{\theta_{{\rm lab}}^{\rm max}} = m_2 / m_1.
\label{maxangle}
\end{equation}
Our main focus is $m_1 = m_\chi$ $>$ $m_2 = m_A$.  For $m_\chi$
somewhat greater than $m_A$, note that the DM scattering angle in the
lab frame is always very forward.

Combining Eqns.~(\ref{Ascaling}), (\ref{scatter}), and (\ref{N}), the
minimum required cross section to capture a DM particle is
\begin{eqnarray}
\sigma_{\chi N}^{\rm min}&=& \frac{m_N^2}{m_A
\left(\frac{\mu(A)}{\mu(N)}\right)^2 \rho L}N_{\rm scat}(m_\chi), 
\label{crosssection} \\
&=&\frac{ -2~\ln\left({v_i}/{v_{\rm{esc}}}\right)}
{\ln\left[1 - 2\frac{m_A  m_\chi}{(m_\chi +
      {m_A})^2}\right]}
\frac{m_N^2}{\left(m_A \left(\frac{\mu(A)}{\mu(N)}\right)^2\rho L \right)},
\nonumber \\
&\underset{m_\chi \gg m_A}{\longrightarrow}& m_\chi \left(\frac{m_N }{m_A}\right)^4 \frac{1}{\rho
  L}\,{\ln\left({v_{i}}/{v_{\rm{esc}}}
\right)}. \nonumber
\end{eqnarray} 
Again, we choose a path length of 0.2 $R_\oplus$, to select about 90\%
of the path lengths in Earth.  Taking this length as a chord through
Earth, the location corresponds to the crust, with an average density
of 3.6 g/cm$^3$, where the most common element is oxygen.  We also
choose an incoming DM velocity of 500 km/s, which effectively selects
the entire thermal distribution.  A slower DM particle is more easily
captured.  These parameters give a required cross section of
\begin{eqnarray}
\sigma_{\chi N}^{\rm min}&=&
\frac{-1.8\times 10^{-33}~\textrm{cm}^2 \left(\frac{\mu(1)}{\mu(16)}\right)^2}
{\ln\left[1 - 2\frac{16~\textrm{GeV} ~m_\chi} {\left(m_\chi +
      16~\textrm{GeV}\right)^{2}}\right]}
\label{sigma}\\
&&\underset{m_\chi \gg m_A}{\longrightarrow}
2.2\times
10^{-37}~\textrm{cm}^2\left(\frac{m_\chi}{\textrm{GeV}}\right).
\label{sigmaapprox}
\end{eqnarray}
Note that we use the unapproximated version, Eq.~(\ref{sigma}), for
our figure, and give the large $m_\chi$ limit in the equations for
demonstrative purposes.  When $m_\chi$ is comparable to $m_A$,
$\sigma_{\chi N}^{\rm min}$ is different from the approximated, large
$m_\chi$ case in an important way.

\begin{figure}[t]
  \includegraphics[width=3.25in,clip=true]{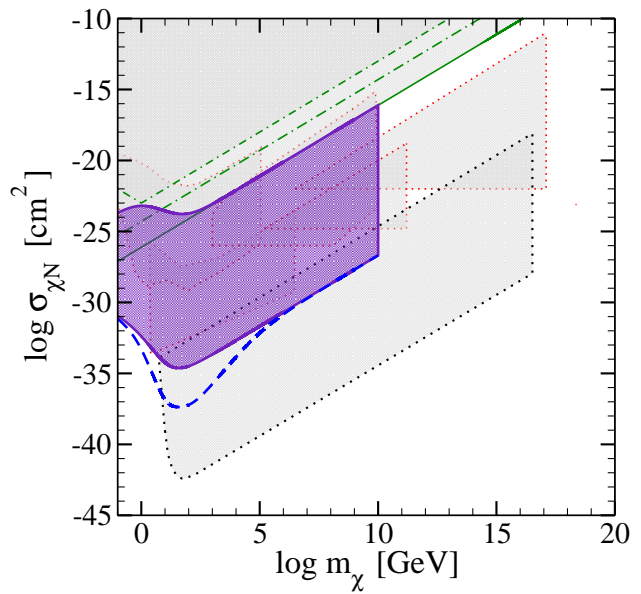}
  \caption{\label{goodplotnew} Inside the heavily-shaded region, dark
  matter annihilations would overheat Earth.  Below the top edge of
  this region, dark matter can drift to Earth's core in a satisfactory
  time.  Above the bottom edge, the capture rate in Earth is nearly
  fully efficient, leading to a heating rate of 3260 TW (above the
  dashed line, capture is only efficient enough to lead to a heating
  rate of $\gtrsim$ 20 TW).  The mass ranges are described in the
  text, and the light-shaded regions are as in Fig.~\ref{regions}.}
\end{figure}

The resulting curve for $\sigma_{\chi N}^{\rm min}$ is shown in
Fig.~\ref{goodplotnew}, as the lower boundary of the heavily-shaded
exclusion region.  The straight section of this constraint is easily
seen from Eq.~(\ref{sigmaapprox}), as the required cross section for
our efficient capture scenario scales as $m_\chi$, due to the large
number of collisions required for stopping, as in
Eq.~(\ref{stopping}).  At lower masses, the curved portion has its
minimum at the mass of the target.  DM masses close to that of the
target can be captured with smaller cross sections because a greater
kinetic energy transfer can occur for each collision.  At very low
masses, much less than the mass of the target, the DM mass dependence
in the logarithm is approximated differently.  In this limit,
$\sigma_{\chi N}$ is $\simeq 10^{-32}~({\rm GeV}/m_\chi)$.  As the DM
mass decreases, it becomes increasingly more difficult for the DM to
lose energy when it strikes a nucleus.  As noted above, the cross
section constraints in the spin-dependent case could be developed, and
would shift the results up by 3 or 4 orders of magnitude.  All of the
other limits that depend on this $A^2$ coherence factor would also
shift accordingly.

The lower edge of the exclusion region is generally rather sharp,
because of these parameters.  For example, consider the case of large
$m_\chi$, where $N_{\rm scat}$ is also large.  If the corresponding
cross section is decreased by a factor $\delta$, so is the number
of scatterings, and by Eq.~(\ref{compounded}), the compounded
fractional kinetic energy loss would only be the 1/$\delta$ root of
that required for capture.  For small cross sections, as usually
considered, the capture efficiency is very low.  To efficiently
produce heat, the minimum cross section must result in $\sim$ 90\% DM
capture.  We stress again that we are not concerned with {\it where}
the DM is captured in Earth, so long as it is.  The probability for
capture can, however, be decreased using Poisson statistics (shown in
Fig.~\ref{goodplotnew} as the dashed line with the accentuated dip at
low masses) to yield just 20 TW of heat flow.  This extension and the
upper edge of the exclusion region are described below.

\section{DM Annihilation and Heating Rates in Earth}
\subsection{Maximal Annihilation and Heating Rates}

Once it is gravitationally captured, DM will continue to scatter with
nuclei in Earth, losing energy until drifting to the core.  Once
there, because of the large cross section, the DM will thermalize with
the nuclei in the core.  The number of DM particles $\mathpzc{N}$ is
governed by the relation between the capture ($\Gamma_C$) and
annihilation ($\Gamma_\mathpzc{A}$) rates~\cite{Griest}:
\begin{equation}
\Gamma_\mathpzc{A} = \frac{1}{2}\mathpzc{A N}^2=\frac{1}{2}\Gamma_C
\tanh^2(t\sqrt{\Gamma_C \mathpzc{A}}).
\label{equilib}
\end{equation}
We neglect the possibility of evaporation~\cite{Gould1} for the
moment, which will affect our results for low $m_\chi$ and
$\sigma_{\chi N}$, as we will explain further below.  The variable $t$
is the age of the system.  $\mathpzc{A}$ is related to the DM
self-annihilation cross section $\sigma_{\chi\chi}$ by
\begin{equation}
\mathpzc{A} = \frac{<\sigma_{\chi \chi} v>}{V_{\rm eff}},
\end{equation}
where $V_{\textrm{eff}}$ is the effective volume of the
system~\cite{Griest}.  For the relevant cross sections considered,
equilibrium between capture and annihilation is generally reached (see
below), so the annihilation rate is
\begin{equation}
\Gamma_\mathpzc{A} = \frac{1}{2} \Gamma_C
\end{equation}

The effective volume is determined by the method of Griest and Seckel
(1988)~\cite{Griest}, which is essentially the volume of the DM
distribution in the core.  The number density of DM is assumed to be
an exponentially decaying function, exp$(-m_\chi \phi/k T)$, like the
Boltzmann distribution of molecules in the atmosphere.  The
temperature of the DM in thermal equilibrium is $T$.  The variable
$\phi$ is the gravitational potential, integrated out to a radius $r$,
written as
\begin{eqnarray}
\phi(r)&=& \int_0^r\frac{G
  M(\tilde{r})}{\tilde{r}^2}d\tilde{r}; \\ 
M(\tilde{r})&=&4\pi\int_0^{\tilde{r}} r^{\prime 2}\rho(r^{\prime})dr^{\prime}.
\end{eqnarray}
The resulting effective volume using the radius of Earth's outer core,
approximately 0.4$R_\oplus$, a temperature of 5000 K, and a density of
9 g/cm$^3$~\cite{Density}, is
\begin{eqnarray}
&V_{\rm eff}&= 4\pi \int_0^{R_{core}} r^2 e^{-\frac{m_\chi \phi}{k
T}}dr.\label{effvolume} \\
&=&1.2\times 10^{25} ~\textrm{cm}^3 \left(\frac{100 ~\textrm{GeV}}{m_\chi}\right)^\frac{3}{2}
\int_0^{0.5\sqrt{\frac{m_\chi}{\textrm{GeV}}}}u^2 e^{-u^2}du
\nonumber \\
&&\underset{m_\chi \gg m_A}{\longrightarrow} 5.3 \times 10^{24} ~{\rm cm}^3
\left(\frac{100 ~\textrm{GeV}}{m_\chi}\right)^\frac{3}{2}.
\end{eqnarray}
For increasing $m_\chi$, the integral (without the prefactor) in the
second line of Eq.~(\ref{effvolume}) quickly reaches an asymptotic
value of about 0.44.

At very large masses, the effective volume for annihilation becomes
very small.  For instance, at $m_\chi$ $\gtrsim$ 10$^{10}$ GeV, the
radius of the effective volume is $\lesssim$ 0.1 km.  With such a
large rate of energy injected in such a small volume, the core
temperature would be increased, requiring a more careful treatment.
However, in Section 5B, we state how a limit on the annihilation cross
section from the unitarity
condition~\cite{Griest:1989wd,Hui:2001wy,Beacom:2006tt} makes us
truncate our bound at $m_\chi$ = 10$^{10}$ GeV, as reflected in
Fig.~\ref{goodplotnew}.  Therefore, this small effective volume is not
a large concern for our exclusion region.

We assume that the DM annihilates into primarily Standard Model
particles, which will deposit nearly all of their energy into Earth's
core (with small corrections due to particle rest masses and the
escape of low-energy neutrinos).  When all of the DM captured is
efficiently annihilated, as specified, the heating rate of Earth is in
equilibrium with the capture rate:
\begin{eqnarray}
\Gamma_{\rm heat} &=& \Gamma_C \times m_\chi = n_\chi\, \sigma_{\rm eff}\, v_\chi \,m_\chi
\nonumber \\
&=& \frac{\rho_\chi}{m_\chi} ~2\pi (0.99 R_\oplus)^2 (270~{\rm km/s})
~m_\chi \\
&=& 3260~ {\rm TW} \nonumber.
\end{eqnarray}
This heat flow is independent of DM mass, since the flux (and capture
rate, when capture is efficient) scales as $1/m_\chi$, while each DM
particle gives up $m_\chi$ in heat when it annihilates.  The value is
much larger than the measured rate of 44 TW we discussed in Section 3.

\subsection{Equilibrium Requirements}

Does the timescale of Earth allow for equilibrium between capture and
annihilation for our scenario?  In order for Eq.~(\ref{equilib}) to be
in the equilibrium limit, $\tanh^2(t\sqrt{\Gamma_C \mathpzc{A}})$ must
be of order unity.  This is true if $t\sqrt{\Gamma_C \mathpzc{A}}$ has
a value of a few or greater.  Since Earth is about 4.5 Gyr old, we
conservatively require that the time taken to reach equilibrium should
be less than about 1 Gyr.  From this, a realistic annihilation cross
section is found.  The condition
\begin{equation}
\tanh^2({t\sqrt{\Gamma_C \mathpzc{A}}}) = 1;~~~ t\sqrt{\Gamma_C \mathpzc{A}}
\simeq \textrm{few}
\end{equation}
allows the relation
\begin{eqnarray}
\frac{<\sigma_{\chi \chi} v>}{V_{\rm eff}} = &\mathpzc{A}& \gtrsim
\frac{(\textrm{few})^2}{\Gamma_C t^2} \\
<\sigma_{\chi \chi} v> &\gtrsim& 10 \frac{V_{\rm eff}}{\Gamma_C t^2}.
\end{eqnarray}
For an efficient capture rate (Eq.~\ref{maxcapture}), the time of 1
Gyr, and the limit of large $m_\chi$, this requires an annihilation
cross section for equilibrium of
\begin{equation}
<\sigma_{\chi \chi} v> ~~\gtrsim~~ 10^{-30} \left(\frac
{\textrm{GeV}}{m_\chi}\right)^{1/2} \textrm{cm}^3/\textrm{s}.
\label{annih}
\end{equation}
Since this required lower bound is much smaller than that of typical
weakly interacting DM particles that are thermal relics
($<\sigma_{\chi \chi} v> \simeq 10^{-26}$ cm$^3$/s~\cite{Bertone}), it
should be easily met.  One expects large scattering cross sections to
be accompanied by large annihilation cross sections, so that even the
possibility of p-wave-only suppression of the annihilation rate should
not be a problem.

For very large masses, the required annihilation cross sectionm, while
small, approaches a quantum mechanical limit.  For example, for
$m_\chi \gtrsim 10^{10}$ GeV the s-wave cross section exceeds the
unitarity bound~\cite{Griest:1989wd,Hui:2001wy,Beacom:2006tt}.  We
note that this may also affect constraints on supermassive DM based on
neutrinos from annihilations~\cite{AHK,Crotty,Albuquerque:2002bj}.  To
be conservative, we therefore do not extend our constraints beyond
this point, though they may still be valid.

The timescale also has to be long enough for DM to drift down to the
core.  If $\sigma_{\chi N}$ is too large, the DM will experience too
many scatterings and will not settle into the core, and thus may not
annihilate efficiently.  Following Starkman et
al. \cite{Starkman:1990nj}, we define the upper edge of our exclusion
region to require a drift time of $\lesssim$ 1 Gyr.  This places a
restriction of
\begin{eqnarray}
\sigma_{\chi N} &\lesssim& 7.7 \times 10^{-20} {\rm cm}^2
\frac{(m_\chi/{\rm GeV})}{A^2(\mu(A)/\mu(N))^2} \nonumber \\
&\lesssim& 7.7 \times 10^{-20} {\rm cm}^2 \frac{m_\chi/{\rm GeV}}{A^4
\left(\frac{m_\chi + m_N}{m_\chi + m_A}\right)^2}\nonumber \\
\sigma_{\chi N} &{\underset{\underset{m_\chi \gg m_A}{\longrightarrow}}{\lesssim}}&
2.5 \times 10^{-23}~\textrm{cm}^{2} \left(\frac{m_{\chi}}{
\textrm{GeV}}\right),
\end{eqnarray}
for a target of iron.  However, a more detailed calculation might
relax this requirement.  For example, Starkman et
al.~\cite{Starkman:1990nj} show that for large values of the capture
cross section and certain other conditions, annihilation may be
efficient enough to occur in a shell, before the DM reaches the core.
This would generally still be subject to our constraint on heat from
DM annihilation, and hence our exclusion region might extend to larger
cross sections than shown.  The two features of this drift line at low
mass occur around the mass of the target and the mass of a nucleon,
due to the various dominances of the mass-dependent term in the
denominator.  The details of the shape of this drift line at low
masses are irrelevant, because the astrophysical constraints already
exclude the corresponding regions.

Aside from drifting to the core, the question of whether heavy DM can
actually get to Earth has been
asked~\cite{Gould:1999je,Lundberg:2004dn}.  The low-velocity tail of
the high-mass DM thermal distribution in the Solar System may be
driven into the Sun by gravitational capture
processes~\cite{Gould:1999je,Lundberg:2004dn}, especially because this
DM's velocity is on the order of the orbital speed of Earth in the
Solar System, which is about 30 km/s. However, this would affect only
a tiny fraction of the full thermal distribution that we require to be
efficiently captured.

\subsection{Annihilation and Heating Efficiencies}
We are not picking a specific model for the annihilation products,
aside from considering only Standard Model particles, which will
deposit their energy in Earth, with the exception of neutrinos.  Our
constraint thus has a very broad applicability.  As noted the
calculated heat flow if DM annihilation is important is 3260 TW, which
is very large compared to our adopted limit on an unconventional
source of 20 TW (or even the whole measured rate of 44 TW).  Typically
then, either DM annihilation heating is overwhelming or it is
negligible, inside or outside of the excluded region.  As shown in
Section 4, the kinetic energy transferred from DM scattering on nuclei
is about 6 orders of magnitude less than the energy from DM
annihilations.  This contribution to Earth's heat is too low to be
relevant for global considerations.  However, it would be interesting
to consider the more localized effect of the kinetic energy deposition
in the atmosphere for very large cross sections.

There are circumstances in which the heating from DM annihilations can
take a more intermediate value, including down to the chosen 20 TW
number.  As explained above, typically the number of scatterings
required to gravitationally capture the DM is very large.  Therefore,
a small decrease in $\sigma_{\chi N}$ and the proportionate change in
the expected number of scatterings means that the compounded kinetic
energy loss is nearly always insufficient.  However, at low $m_\chi$,
the number is small enough that upward fluctuations relative to the
expected number can lead to capture.  If $N_{\rm scat}$ collisions
typically lead to efficient capture for a cross section $\sigma_{\chi
N}^{\rm min}$, as defined above, a new and smaller $N$ may be defined
by the condition that the Poisson probability Prob($N \geq N_{\rm
scat}$) = 20 TW / 3260 TW = 1/163.  With this $N$, and its
proportionately smaller $\sigma_{\chi N}$, upward Poisson fluctuations
in the number of scatterings lead to efficient capture for a fraction
1/163 of the incoming flux.  Note that this small capture fraction is
not just the low-velocity tail of the DM thermal distribution, since
we have defined these conditions for the highest incoming velocities,
$v_i$ = 500 km/s.

The resulting constraint on $\sigma_{\chi N}$ is shown by the dashed
line that dips below the main excluded region in
Fig.~\ref{goodplotnew}.  The enhanced valley around 16 GeV again
arises from the ease of capture when the DM mass is near the target
mass.  Note that for each mass the required $<\sigma_{\chi \chi} v>$
is increased by the same factor that decreased the original required
$\sigma_{\chi N}^{\rm min}$.  Since most of this exclusion region is
already covered by underground detectors, its details may not be so
important.

For low DM masses, evaporative losses of DM from the core due to
upscattering by energetic iron nuclei may be relevant~\cite{Gould1}.
Simple kinematic estimates show that DM masses below $\simeq$ 5 GeV
might be affected.  However, this is only potentially important if
$\sigma_{\chi N}$ is small enough that scattering is very rare --
since otherwise any upscattered DM will immediately downscatter.  From
the considerations above about Poisson fluctuations in the number of
scatterings, we expect that this should only be relevant between the
dark-shaded region and the dashed line.

\section{Discussion and Conclusions}
\subsection{Principal Results}

As summarized in Fig.~\ref{regions}, while very large DM-nucleon
scattering cross sections are excluded by astrophysical
considerations, and small cross sections are excluded by underground
direct DM detection experiments, there is a substantial window in
between that has proven very difficult to test, despite much
effort~\cite{Starkman:1990nj,Rich:1987st,Wandelt,Erickcek:2007jv,Gould1,Gould2,Gould3,
Gould4,Gould7,KSW,Uranus,Kawasaki:1991eu,AHK,Desai:2004pq,Bottino:1994xp,
Achterberg:2006jf,Banks:2005hc,Enqvist:2001jd,McGuire:1994pq,
Shirk:1978,Snowden-Ifft:1990,Anderson:1995}.  High-altitude
experiments have excluded only parts of this window.  In this window,
DM will be efficiently captured by Earth.  We point out that the
subsequent self-annihilations of DM in Earth's core would lead to an
enormous heating rate of 3260 TW, compared to the geologically
measured value of 44 TW.

We show that the conditions for efficient capture, annihilation, and
heating are all quite generally met, leading to an exclusion of
$\sigma_{\chi N}$ over about ten orders of magnitude, which closes the
window on strongly interacting DM between the astrophysical and direct
detection constraints.  These new constraints apply over a very large
mass range, as shown in Fig.~\ref{goodplotnew}.  We have been quite
conservative, and so very likely an even larger region is excluded.
These results establish that DM interactions with nucleons are bounded
from above by the underground experiments, and therefore that these
interactions must be truly weak, as commonly assumed.  This means that
direct detection experiments are looking in the correct $\sigma_{\chi
N}$ range when sited underground and motivates further theoretical
study of weakly interacting
DM~\cite{Vergados:2006sy,Anchordoqui:2005is,Hooper:2006wv,Baer:2005zc}.
Furthermore, it means that DM-nucleon scattering cannot have any
measurable effects in astrophysics and cosmology, and this has many
implications for models with strongly or moderately interacting
DM~\cite{Banks:2005hc,Enqvist:2001jd,Zaharijas:2004jv,Khlopov} and other
astrophysical constraints on the DM-nucleon interaction cross
section~\cite{Furlanetto:2001tw,Chuzhoy:2004bc}.  This exclusion
region also completely covers the cross section range in which
strongly interacting dark matter might bind to
nuclei~\cite{Javorsekandmore}.

To evade our constraints, extreme assumptions would be required: that
DM is not its own antiparticle, or that there is a large ( $\gtrsim
163$) particle-antiparticle asymmetry from DM, or that DM
self-annihilations proceed only to purely sterile non-Standard Model
particles, at the level of $\gtrsim$ 100:1.  (Although in
Ref.~\cite{Beacom:2006tt} such a large branching ratio to neutrinos
was considered, it was emphasized that this was used only to set the
most conservative bound on the DM annihilation cross section, and not
to be representative of a realistic model.)  While our constraints are
based on Earth's measured heat flow, it is important to emphasize that
DM capture and annihilation generally cannot contribute measurably
without being overwhelming, and hence are excluded.  Thus, in the
ongoing debate over the unknown sources of Earth's heat flow, it seems
that DM can play no role.  The most important next step in refining
our understanding of the known generators of the measured 44 TW will
come from isolating the contribution from uranium and thorium decays
by measuring the corresponding neutrino
fluxes~\cite{Araki:2005qa,Fiorentini:2005ma,Fiorentini1,Fields,Krauss:1983zn,
Hochmuth:2005nh,Dye:2006gx,Fogli:2006zm,Miramonti:2006kp,Rothschild:1997dd,Raghavan:1997gw}.

\subsection{Comparison to Other Planets}

Starkman et al.~\cite{Starkman:1990nj} calculated the efficient Earth
capture line for DM, but only had model-dependent results.  We have
now considered the consequences of annihilation in Earth, and have
shown that it gives a model-independent constraint.  Other planets
have been discussed, such as Jupiter and Uranus
\cite{KSW,Fukugita:1988,Dimopoulos:1989hk,Kawasaki:1991eu,Abbas:1996kk,Uranus,Zaharijas:2004jv,Farrar:2005zd},
but Earth is the best laboratory.  It is the best understood planet,
with internal heat flow data measured directly underground, from many
locations, and Earth's composition and density profile are well
known~\cite{TextBook,Pollack,BlackBook,vFrese,Panero,OrangeBook,Density}.
Importantly, the relative excess heat due to DM annihilation would be
much greater for Earth than the Jovian planets.

What about other planets?  The maximal heating rate due to DM scales
with surface area, and can be compared with the internal heating rates
estimated from infrared data~\cite{AstroBook}.  If a constraint can be
set, the minimum cross section $\sigma_{\chi N}^{\rm min}$ that can be
probed scales with the planet's column density as $(nL)^{-1}$ (see
Eq.~(\ref{N})) up to nontrivial corrections for composition (see
Eq.~(\ref{Ascaling})).  Note that the column density $nL$ is
proportional to the surface gravity $\sim$ $GM/R^2$, which varies
little between the planets, as noted in Table II~\cite{AstroBook}.
Due to its known (and heavy) composition and well-measured (and low)
internal heat, the strongest and most reliable constraints will be
obtained considering Earth.  As an interesting aside, it may then be
unlikely that heating by DM could play a significant role in
explaining the apparent overheating of some extra-solar planets (``hot
Jupiters'')~\cite{Gaudi:2004rk,Bakos:2006kb,Burrows:2006pf,Fabrycky:2007pw,Southworth:2007kc}.

\begin{table}
\caption{Comparison of potential dark matter constraints using various planets~\cite{AstroBook}.
A greater difference between dark matter and internal heating rates
give greater certainty.  The minimum cross section probed scales
roughly with the surface gravity.  Earth is the best for setting
reliable and strong constraints.}
\begin{center}
  \begin{tabular}{|c|c|c|c|}
    	\hline 
    	Planet &  DM Max. Heating & Internal Heat & Surface Gravity\\
	 & (TW) & (TW) & (units of $\oplus$)\\
    	\hline
	\hline
    	Earth  & 3.3 $\times~10^3$ & 0.044 $\times~10^3$ & 1.00\\
    	\hline 
    	\hline
	Jupiter & 420 $\times~10^3$ & 400 $\times~10^{3}$ & 2.74\\
    	\hline 
    	Saturn & 290 $\times~10^3$ & 200 $\times~10^{3}$ & 1.17\\
    	\hline 
    	Uranus & 53 $\times~10^3$  & $\leq ~10^{3}$ & 0.94\\
    	\hline 
    	Neptune & 50 $\times~10^3$ & 3 $\times~10^{3}$ & 1.15\\
    	\hline
  \end{tabular}
\end{center}
\end{table}
\subsection{Future Directions}

While it appears that the window on strongly interacting DM is now
closed over a huge mass range, more detailed analyses are needed in
order to be absolutely certain.  Our calculations are conservative,
and the true excluded region is likely to be larger.  It would be
especially valuable to have new analyses of the astrophysical limits
and the underground detector constraints.  This would give greater
certainty that no sliver of the window is still open.  For DM masses
in the range 1 GeV to 10$^{10}$ GeV, the upper limits on the DM
scattering cross section with nucleons from CDMS and other underground
experiments have been shown to be true upper limits.  Thus the DM does
indeed appear to be very weakly interacting, and it will be
challenging to detect it.

\begin{acknowledgments}
We are grateful for discussions and advice from Nicole Bell, Eric
Braaten, Matt Kistler, Jordi Miralda-Escud\'{e}, Jerry Newsom, Wendy
Panero, Gary Steigman, Louie Strigari, Ralph von Frese, Terry Walker,
and especially Laura Baudis, Andy Gould, Stefano Profumo, and Hasan
Y\"uksel.  GDM was supported by the Department of Energy Grant
DE-FG02-91ER40690, JFB by the National Science Foundation CAREER grant
PHY-0547102, and GB during part of this project by the Helmholtz
Association of National Research Centres under project VH-NG-006.
\end{acknowledgments}


\end{document}